\documentclass[%
 aps,
 amsmath,amssymb,
preprint,%
]{revtex4-2}

\usepackage{graphicx}
\usepackage{dcolumn}
\usepackage{bm}

\usepackage[utf8]{inputenc}
\usepackage[T1]{fontenc}
\usepackage{mathptmx}
\usepackage{etoolbox}
\usepackage{color}

\makeatletter
\def\@email#1#2{%
 \endgroup
 \patchcmd{\titleblock@produce}
  {\frontmatter@RRAPformat}
  {\frontmatter@RRAPformat{\produce@RRAP{*#1\href{mailto:#2}{#2}}}\frontmatter@RRAPformat}
  {}{}
}%
\makeatother
\begin{document}


\title[Optically excited spin dynamics of thermally metastable skyrmions in Fe$_{0.75}$Co$_{0.25}$Si]{Optically excited spin dynamics of thermally metastable skyrmions in Fe$_{0.75}$Co$_{0.25}$Si}
\author{J. Kalin$^*$}\email{jantje.kalin@ptb.de}
\affiliation{Physikalisch-Technische Bundesanstalt, 38116 Braunschweig, Germany}

\author{S. Sievers}%
\affiliation{Physikalisch-Technische Bundesanstalt, 38116 Braunschweig, Germany}

\author{F. Garc\'{i}a-S\'{a}nchez}%
\affiliation{Departamento de F\'{i}sica Aplicada, Universidad de Salamanca, 37008 Salamanca, Spain}

\author{A. Bauer}
\affiliation{Physik-Department, Technische Universität München, 85748 Garching, Germany}
\affiliation{Zentrum für QuantumEngineering (ZQE), Technische Universität München, 85748 Garching, Germany}

\author{H. Füser}
\affiliation{Physikalisch-Technische Bundesanstalt, 38116 Braunschweig, Germany}

\author{H. W. Schumacher}
\affiliation{Physikalisch-Technische Bundesanstalt, 38116 Braunschweig, Germany}

\author{C. Pfleiderer}
\affiliation{Physik-Department, Technische Universität München, 85748 Garching, Germany}
\affiliation{Zentrum für QuantumEngineering (ZQE), Technische Universität München, 85748 Garching, Germany}
\affiliation{Munich Center for Quantum Science and Technology (MCQST), Technische Universität München, 85748 Garching, Germany}

\author{M. Bieler}
\affiliation{Physikalisch-Technische Bundesanstalt, 38116 Braunschweig, Germany}

\date{\today}

\begin{abstract}
We investigate the microwave spin excitations of the cubic chiral magnet Fe$_{0.75}$Co$_{0.25}$Si as driven by the thermal modulation of magnetic interactions via laser heating and probed by time-resolved measurements of the magneto-optical Kerr effect. Focusing on the topologically nontrivial skyrmion lattice state, the dynamic properties in thermodynamic equilibrium are compared with those of a metastable state prepared by means of rapid field cooling. In both cases, we find precessional and exponential contributions to the dynamic response, characteristic of a breathing mode and energy dissipation, respectively. When taking into account the universal scaling as a function of temperature, the precession frequencies in the equilibrium and metastable skyrmion state are in excellent quantitative agreement. This finding highlights that skyrmion states far from thermal equilibrium promise great flexibility, for instance with respect to temperature and field scales, both for possible microwave applications and the study of fundamental properties.
\end{abstract}
\maketitle
Magnetic skyrmions are nanoscale spin whirls that show inherent robustness against external perturbation due to their special topology and have nontrivial magnetoelectrical properties, such as their intriguing collective spin dynamics in the GHz-frequency region \cite{mochizuki2012spin, onose2012observation, schwarze2015universal, okamura2013microwave}. 
Due to that skyrmion hosting materials are considered a potential candidate for the development of future microwave- and magnonics-related applications \cite{garst2017collective, mochizuki2013magnetoelectric, finocchio2015skyrmion,xing2020magnetic}.
This has triggered intensive research on skyrmion dynamics in cubic chiral magnets \cite{lonsky2020dynamic, mochizuki2012spin,onose2012observation, okamura2013microwave, schwarze2015universal, ogawa2015ultrafast, seki2020propagation, pollath2019ferromagnetic, aqeel2021microwave, seaberg2017nanosecond, seaberg2021spontaneous, takagi2021hybridized, seki2020propagation}. 
In comparison to many skyrmion hosting materials based on complex multilayers, the skyrmions in cubic chiral magnets form a regular hexagonal lattice \cite{yu2010real}, which allows to study fundamental properties of collective skyrmion dynamics.
In the pioneering theoretical work by Mochizuki the dominating skyrmion eigenmodes (i.e., breathing and gyration modes) and their selection rules were proposed and soon after experimentally verified for several materials \cite{schwarze2015universal, onose2012observation,ogawa2015ultrafast, padmanabhan2019optically}.
Despite these efforts, one important obstacle for the transition towards successful skyrmion devices still remains: The skyrmions in cubic chiral magnets exist only in a small temperature and magnetic field region, difficult to reach in applications. 
The observation of a thermodynamically metastable skyrmion state in cubic chiral magnets \cite{munzer2010skyrmion,bauer2016history,bauer2018skyrmion,oike2016interplay,berruto2018laser, wild2017entropy,ritz2013giant, okamura2016transition, yu2018aggregation, bannenberg2016extended} over large parts of the magnetic phase diagram with long lifetime \cite{wild2017entropy} paves a new pathway for the realization of microwave- and magnonics-related applications.
However, so far, little is known on how the skyrmion dynamics develop outside the parameter regime of the equilibrium state in the non-thermal limit and how they compare to the equilibrium dynamics.

In this letter, we explore the skyrmion dynamics in the chiral B20 magnet Fe$_{\text{1-x}}$Co$_{\text{x}}$Si in thermodynamic equilibrium and in the metastable state using time-resolved magneto-optical Kerr effect (TR-MOKE) measurements.
In contrast to the microwave magnetic field \cite{schwarze2015universal, mochizuki2012spin,onose2012observation} and magneto-optical excitation \cite{ogawa2015ultrafast} applied so far to study the dynamics of B20 magnets, we use a thermal excitation mechanism by laser heating with linearly polarized light \cite{comment1}. The spin dynamics are triggered by the thermal modulation of the effective field 
due to the optically induced temperature change \cite{van2002all}. As the dynamics are driven by an indirect coupling between photons and the spin system \cite{beaurepaire1996ultrafast}, this technique enables the time-resolved characterization not only of precession modes, but also of energy dissipation processes after laser excitation. 
We study and compare the magnetic field and temperature dependence of these different dynamical contributions in the thermodynamic equilibrium and metastable skyrmion state of Fe$_{\text{1-x}}$Co$_{\text{x}}$Si for the first time.
While our study shows a higher potential of the metastable skyrmion state for possible microwave applications as compared to the equilibrium state, it also demonstrates that the metastable skyrmion state is well suited to investigate generic properties of skyrmion dynamics.

\begin{figure}
\includegraphics{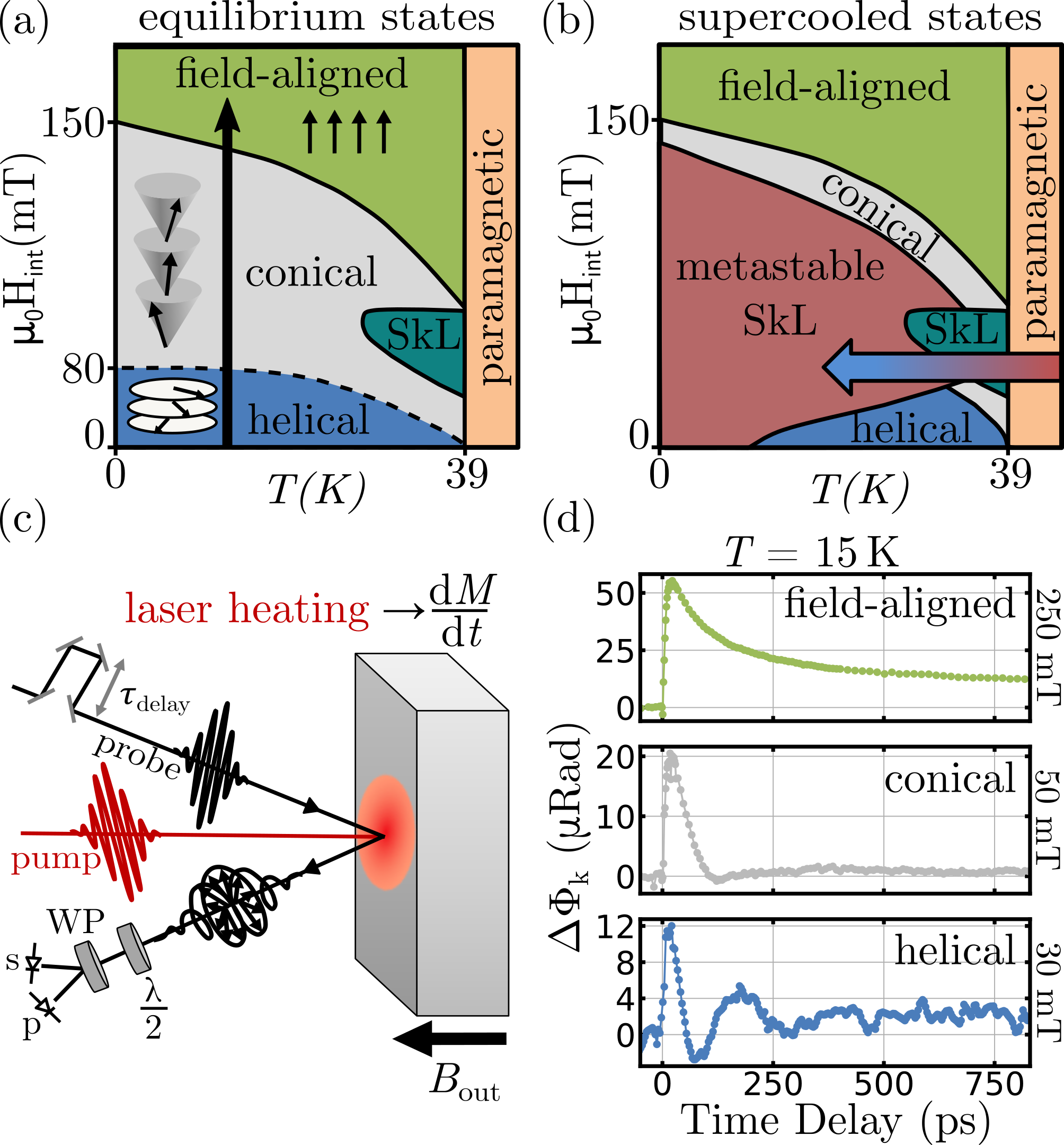}
\caption{\label{fig:Figure1} Sketch of the magnetic phase diagram of Fe$_{\text{0.75}}$Co$_{\text{0.25}}$Si in thermal equilibrium (a) and under supercooling for rapid cooldowns through the skyrmion pocket (b) with internal magnetic field values $\mu_0 H_{int}$ taken from Ref. \cite{bauer2016history}. The dashed line in (a) indicates that the helical phase is only present for zero-field cooling. (c) Schematic optical setup for excitation and detection of the spin dynamics by TR-MOKE. (d) Typical TR-MOKE signal for the field-aligned (green), conical (gray) and helical (blue) phase at $15\,$K. The black arrow in (a) shows the change of the magnetic field during these measurements.  
 }
\end{figure}

The sample used in this study is a ${2\times 2 \times 0.5\,\text{mm}^3}$ large Fe$_{0.75}$Co$_{0.25}$Si crystal with all faces perpendicular to the magnetically easy axes $<$100$>$ and a magnetic ordering temperature of ${T_\text{c} = 39\,\text{K}}$ \cite{bauer2016history}. Details about the magnetic properties and crystal growth are provided in Ref. \cite{bauer2016history}. In thermal equilibrium, Fe$_{\text{1-x}}$Co$_{\text{x}}$Si shows the generic magnetic phase diagram of B20 magnets \cite{bauer2016generic, muhlbauer2009skyrmion, munzer2010skyrmion,seki2012observation,adams2012long}, schematically depicted in Fig. \ref{fig:Figure1}(a). At high temperatures Fe$_{\text{1-x}}$Co$_{\text{x}}$Si is in a paramagnetic state. Below $T_\text{c}$ the helical phase is observed for fields smaller than the critical field $H_{\text{c1}}$. This phase is characterized by long-wavelength helices, which are aligned with the magnetically easy axes $<$100$>$ given by the cubic anisotropy. 
At the phase boundary the spins tilt towards the magnetic field direction and form the conical phase. 
Above the critical field $H_{\text{c2}}$ the spins align collinearly in the field-aligned phase. 
Note here, that the helical phase of Fe$_{\text{1-x}}$Co$_{\text{x}}$Si is only observed for zero-field cooling (cooldown at $B=0\,T$) \cite{bauer2016history}, characteristic for doped compounds \cite{bauer2018skyrmion}. After a cooldown under magnetic fields, e.g., high-field cooling with $H>H_{\text{c2}}$, the conical phase extends to zero magnetic fields \cite{bauer2016history}.
Most prominent in the magnetic phase diagram is the small area just below $T_\text{c}$ for intermediate fields where the periodic magnetic skyrmion lattice (SkL) with hexagonal symmetry \cite{yu2010real}, the so called skyrmion pocket, is observed. 
Beyond that, skyrmions in Fe$_{\text{1-x}}$Co$_{\text{x}}$Si can also be generated outside the parameter regime of their equilibrium state by supercooling \cite{munzer2010skyrmion,bauer2016history,bauer2018skyrmion, wild2017entropy,ritz2013giant, okamura2016transition, yu2018aggregation, bannenberg2016extended}. In this case the phase transition between the skyrmion pocket and equilibrium conical phase is suppressed. As a consequence the skyrmion lattice evolves as thermodynamically metastable state (MSkL), which extends over large parts of the magnetic phase diagram (see Fig. \ref{fig:Figure1}(b)). To that end the material is cooled down rapidly from temperatures well above $T_\text{c}$ to the desired temperature, while applying a magnetic field which allows to cross the phase pocket of the equilibrium SkL \cite{bauer2018skyrmion}. This cooldown procedure is called rapid field cooling and is used in this study to characterize skyrmion dynamics in non-thermal equilibrium.

The experimental pump-probe setup for TR-MOKE is depicted in Fig. \ref{fig:Figure1}(c). 
Laser pulses of about $150\,$fs duration with photon wavelength of $800\,$nm and a repetition rate of $76\,$MHz were generated by a Ti:Sapphire oscillator. The linearly polarized laser light was then split into a pump and probe beam.
 The pump beam ($60\,\mu\text{m} \text{ beam radius}$) thermally excites the Fe$_{\text{0.75}}$Co$_{\text{0.25}}$Si sample by laser heating and triggers the magnetization dynamic. Next to steady state heating, which we take into account when estimating the so called base temperature $T_0$, the laser heating leads to a temperature change on a picosecond time-scale. The transient temperature evolution can be described phenomenologically by the Three-Temperature-Model \cite{beaurepaire1996ultrafast}. The induced change of spin temperature causes a magnetization change and leads to the excitation of spin dynamics. For an absorbed laser fluence of $F= 2 \,\mu\text{J}/\text{cm}^2$ and a laser penetration depth of $\xi = 30\,\text{nm}$, we estimate that the maximum transient spin temperature $T_{\text{s,max}}$ in our experiment is $30 \pm 4\,\text{K}$,  $34 \pm 4\,\text{K}$ and $39 \pm 4\,\text{K}$ for a base temperature $T_0$ of $10\,\text{K}$,  $20\,\text{K}$ and $30\,\text{K}$, respectively.

 The magnetization dynamics are detected by analyzing the polarization state of the probe beam ($30\,\mu\text{m} \text{ beam radius}$), which rotates by a Kerr angle $\phi_\text{k}$ upon reflection from the sample surface. For this purpose, we use a polarization bridge consisting of a half-wave plate ($\lambda/2$), Wollaston prism (WP) and a pair of balanced photodiodes. To enhance the detection sensitivity, the pump beam is mechanically amplitude modulated at $3\,$kHz. The TR-MOKE signal is extracted from the balanced photodiode voltage by subsequent lock-in demodulation. Accordingly, we measure the change of the Kerr angle, $\Delta \phi_\text{k}$, with and without laser excitation, which is proportional to a magnetization change. Time-resolved measurements are realized by adjusting the delay between pump and probe beam, so that the change in magnetization can be measured as a function of time after laser excitation with sub-picosecond resolution.
The two laser beams hit the sample at almost normal incidence and consequently the TR-MOKE signal is mostly sensitive to the variation of the out-of-plane magnetization component \cite{you1998generalized}. In the experiment, the applied magnetic field and sample temperature were controlled using a cryostat with superconducting magnet, where the applied magnetic field is always oriented out of the sample plane in $<$100$>$ crystal direction.  

First, we characterize the equilibrium states at low temperatures. Typical TR-MOKE signals for the field-aligned, conical and helical phase at $15\,$K are shown in Fig. \ref{fig:Figure1}(d). The measurements are performed subsequent to zero-field cooling in a field scan going from zero to high magnetic fields, which is shown schematically by the black arrow in Fig. \ref{fig:Figure1}(a).
In the field-aligned and conical phase the TR-MOKE signal rises after the thermal excitation by the optical pump beam at zero time and reaches its maximum after approximately $20\,$ps. In the helical phase, the maximum occurs at about $15\,$ps. 
We attribute the initial rise of the TR-MOKE signal to the thermally induced reduction of the sample magnetization (demagnetization) by laser heating. It is much slower than what is observed for 3d ferromagnets \cite{beaurepaire1996ultrafast}.
After the maximum signal is reached the TR-MOKE signal decreases corresponding to a magnetization recovery. In the field-aligned phase we observe a monotonous decay of the TR-MOKE signal, which does not reach its initial state (at $t < 0$) within ${800\,\text{ps}}$. Hereby, the decay does not change significantly with the applied magnetic field (not shown).
In comparison, in the conical phase the TR-MOKE signal decays on a much faster timescale with a small overshoot to negative values. This dynamic gets faster with the applied magnetic field and the overshoot rises for smaller magnetic field values (not shown).
In the helical phase the rapid rise of the TR-MOKE signal is followed by a highly damped GHz oscillation. This behavior is a clear signature of the excitation of collective precessional out-of-plane spin dynamics and thus of helical eigenmodes, the so called helimagnons as already observed for Fe$_{0.8}$Co$_{0.2}$Si \cite{Koralek2012}. 
In our measurements the field-aligned and the conical phases do not reveal an oscillatory signal as typical fingerprint of collective precessional dynamics. Note however, that the presence of such collective dynamics cannot be ruled out since they might dominantly only occur in the sample plane to which our setup is insensitive. 
Despite of this, the equilibrium states can be unambiguously identified using characteristic signatures within the TR-MOKE signal.
\begin{figure}
\includegraphics{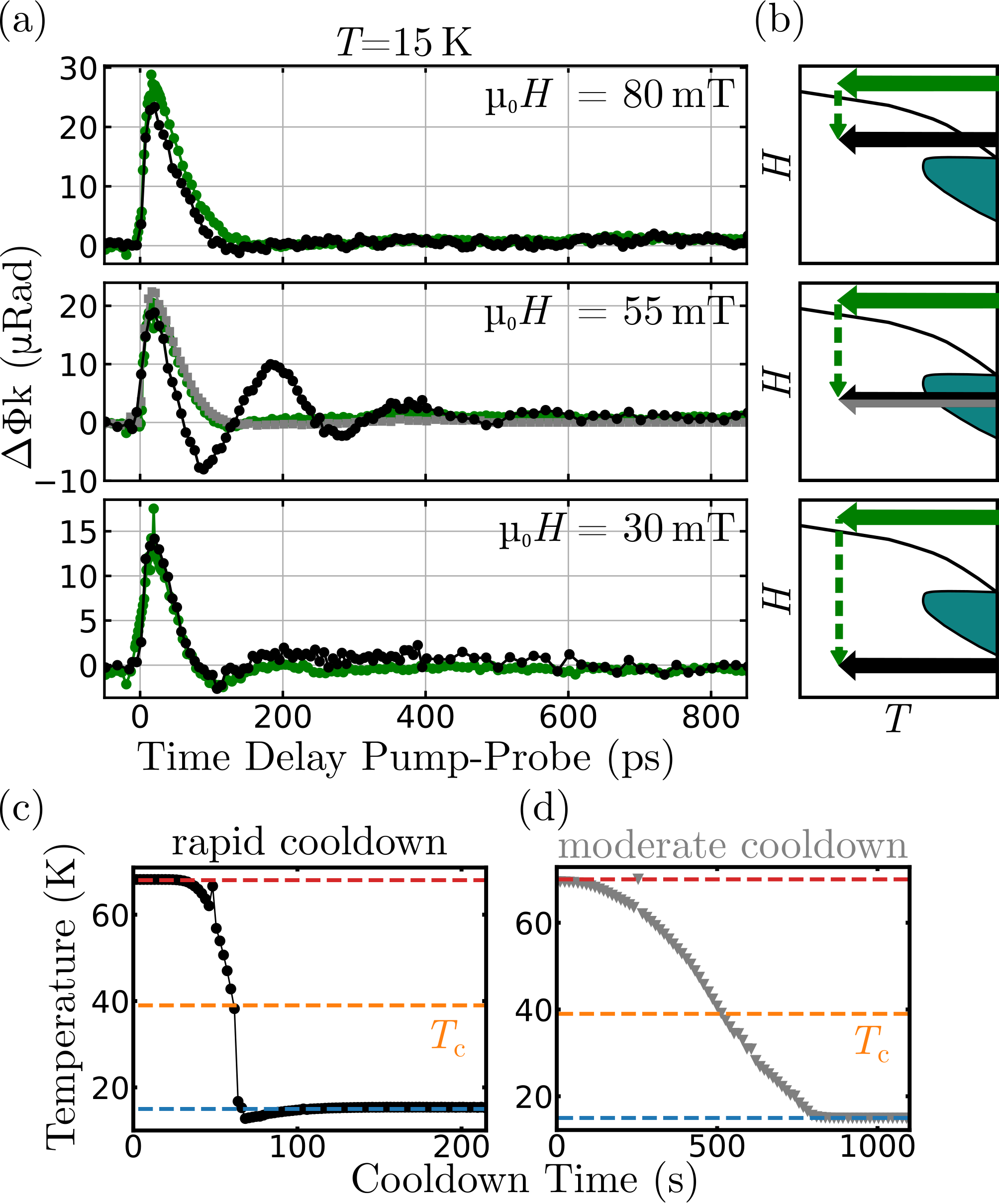}
\caption{\label{fig:Figure2}(a)	TR-MOKE measurements subsequent to rapid field cooling (black curves), high-field cooling (green curves) and field cooling (gray curve) for $80\,$mT (upper panel), $55\,$mT (center panel) and $30\,$mT (lower panel). (b)	The solid arrows in the sketched magnetic phase diagram illustrate the change of temperature during the rapid field cooling (black), high-field cooling (green) and field cooling (gray), respectively. The dashed green arrow indicates the change of magnetic field after the high-field cooling.
Temperature evolution of a sensor in sample proximity during rapid (c) and moderate (d) cooldown. The sample is cooled down from T = $70\,$K (red line)  > $T_\text{c}$ (orange line) to $15\,$K (blue line).
}
\end{figure}

We now explore the magnetization dynamics of Fe$_{0.75}$Co$_{0.25}$Si after rapid field cooling to ${15 \,\text{K}}$.
Taking the phase transition fields from ac susceptibility measurements \cite{bauer2016history}, we apply three characteristic magnetic field values: ${H = 30\,\text{mT}}$ (below SkL), ${H = 55\,\text{mT}}$ (center SkL), and ${H = 80\,\text{mT}}$ (above SkL), see Fig. \ref{fig:Figure2}(a-b).
The sample is cooled via the helium bath provided by the cryostat from ${70 \text{ K} > T_\text{c}}$ to $15\,$K and the cooldown is realized by a sudden increase of the helium flow in the sample chamber. The temperature evolution of a sensor placed in proximity to the sample in the cryostat for rapid cooldowns is shown in Fig. \ref{fig:Figure2}(c). 
At zero time the cooling is initialized. The temperature sensor shows an abrupt cooldown within $30\,$s and settles soon after to the temperature setpoint of $15\,$K. We take this as an estimation of the time-dependent sample temperature.
In the upper panel in Fig. \ref{fig:Figure2}(a) we show the TR-MOKE measurement subsequent to rapid cooldowns at magnetic fields above the field span of the equilibrium SkL exemplary for $80\,$mT.
The TR-MOKE signal shows the characteristic of the conical phase measured after high-field cooling (green curves and also compare with Fig. \ref{fig:Figure2}(b)). The same we observe for rapid cooldowns at magnetic fields below the field span of the equilibrium SkL, shown for $30\,$mT in the lower panel in Fig. \ref{fig:Figure2}(a). Recall here that the helical phase, which is present for zero-field cooling in this field range, is not recovered under field cooling.
When we cross the equilibrium SkL during the rapid field cooling (see center panel of Fig. \ref{fig:Figure2}(a)), we observe different dynamics than for the equilibrium states at the same magnetic field and temperature. The TR-MOKE signal shows a damped GHz oscillation, which settles to the initial value ($t < 0$) after approximately $600\,\text{ps}$. We assign this behavior to the excitation of collective spin dynamics in the MSkL, which we stabilized via the rapid field cooling.
With that, we demonstrate here the first experimental observation of collective skyrmion dynamics far from thermal equilibrium in Fe$_{0.75}$Co$_{0.25}$Si.
It is worth noticing that even though we heat the sample locally by laser excitation during the experiment, the MSkL remains stable on timescales relevant for the characterization. Most likely this is because we do not increase the transient spin temperature by laser heating above the temperature range where the MSkL is stable (see Fig.\ref{fig:Figure1}(b)) according to our temperature estimate. 
In contrast to previous works \cite{munzer2010skyrmion, bauer2016history} a moderate cooldown through the skyrmion pocket (see Fig. \ref{fig:Figure2}(d)) is not sufficient to stabilize the MSkL in our experiment. We observe again the conical state, shown as the gray curve in Fig. \ref{fig:Figure2}(a). Possibly, temperature gradients across the Fe$_{0.75}$Co$_{0.25}$Si sample due to local laser heating during the field cooling process promote the unwinding of the equilibrium skyrmion lattice (skyrmion collapse) in our experiments, requiring higher cooling rates for the stabilization of the MSkL.

\begin{figure}[!]
\includegraphics{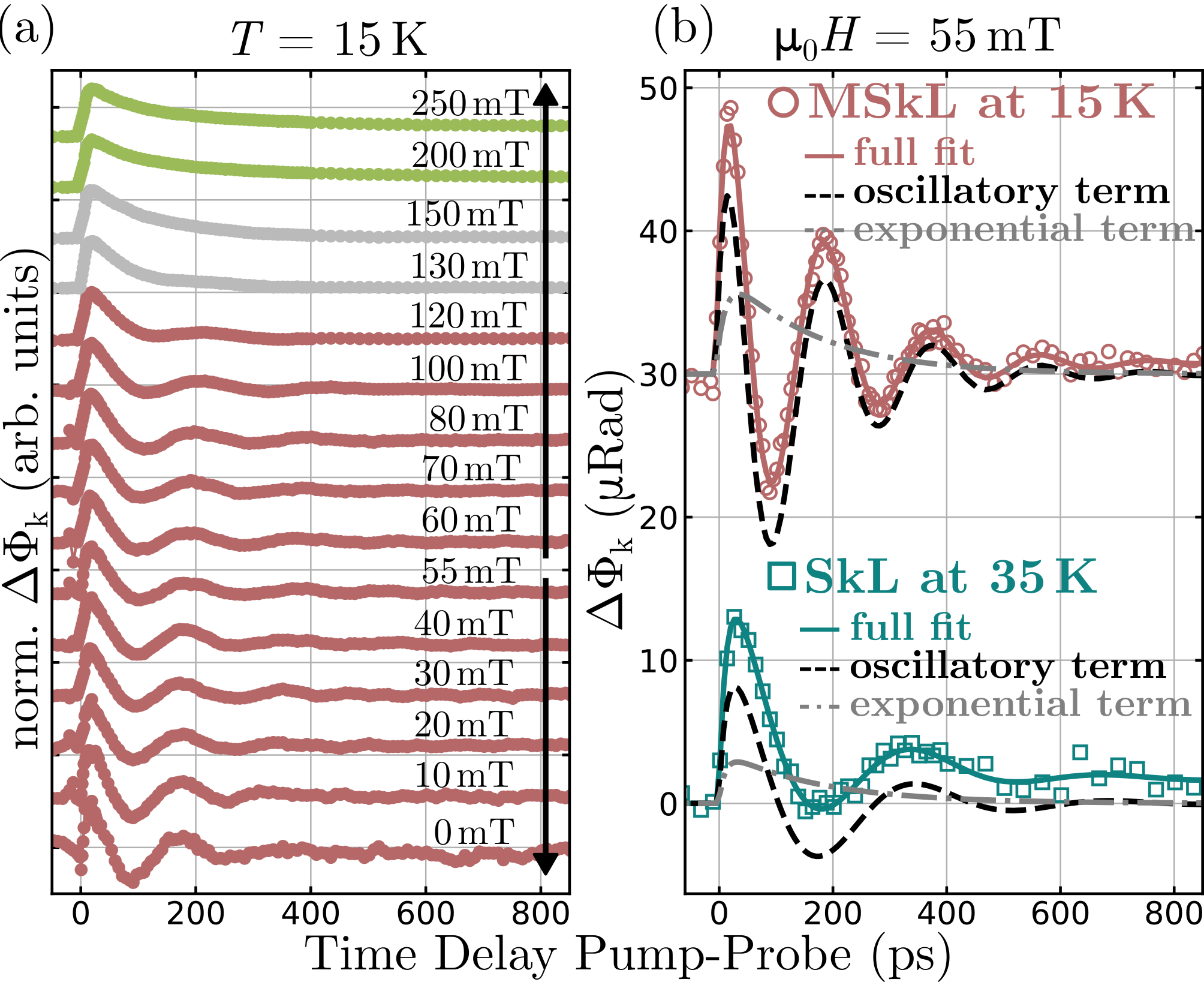}
\caption{\label{fig:Figure3} (a) TR-MOKE measurements at $15\,$K for different magnetic fields after rapid field cooling at $55\,$mT. The measurements were collected in field sweeps with increasing (vertical arrow upwards) and decreasing (vertical arrow downwards) fields.
(b)	Comparison of the skyrmion dynamics in the MSkL at $15\,$K (red circle) and the equilibrium SkL at $35\,\text{K}$ (turquoise square) at $55\,\text{mT}$. The solid lines show the fit with the phenomenological model and the dashed lines represent the exponential and oscillatory component.  The data of the MSkL is offset by $30\, \mu$Rad for clarity.
}
\end{figure}
Having established the existence of the MSkL in our experiment, we further investigate its dynamics versus magnetic field at $15\,$K, see Fig. \ref{fig:Figure3}(a). The sample is rapidly cooled down at $55\,$mT from ${T >T_\text{c}}$ to $15\,$K. The TR-MOKE measurements are recorded in field sweeps for increasing and decreasing fields. Before each scan the cooldown procedure is repeated. We observe that for magnetic fields ranging from $0$ to $120\,$mT, the precessional dynamics associated with the MSkL persist. This is true even outside the magnetic field range where the equilibrium SkL exists.
For fields larger than $120\,$mT we observe again the TR-MOKE characteristic of the conical phase.
This can be explained by the unwinding of the skyrmions for a critical magnetic field value and the transition to the equilibrium state. Above H$_{\text{c2}}$ we enter the field-aligned phase, as indicated by the slow remagnetization, compare with Fig. \ref{fig:Figure1}(d).

Next, we address the question how the skyrmion dynamics in the MSkL compare to the one in the equilibrium SkL.
To this end we plot in Fig. \ref{fig:Figure3}(b) the TR-MOKE signal obtained from the MSkL (${15\,\text{K}, 55\,\text{mT}}$) and from the equilibrium SkL (${35\,\text{K}, 55\,\text{mT}}$). As in the MSkL we also observe in the SkL an oscillation of the TR-MOKE signal resulting from a precessional mode, but with smaller amplitude and lower frequency.
For a quantitative comparison of the magnetization dynamics, we use a phenomenological model which separates oscillatory and exponential components.
Within this framework the TR-MOKE signal $\Delta \phi_\text{k}$ is given by 
\begin{equation}
\Delta \phi_\text{k} = \left(e^{-t/\tau_\text{k}}(A+B \sin(2\pi f_\text{p} t+\delta))+C\right)  \left(1-e^{-t/\tau_{\text{rise}}}\right)\Theta(t) \label{eq:phenomenologicalModel}
\end{equation}
The energy dissipation after laser excitation is modeled by the exponential term with amplitude $A$. The rise-time $\tau_{\text{rise}}$ is attributed to the demagnetization and the decay time $\tau_\text{k}$ to the magnetization recovery. 
The precessional spin dynamics are described by the oscillatory component with frequency $f_\text{p}$ and amplitude $B$. We note here that phase offsets $\delta$ are required to fit the early dynamics of the TR-MOKE data properly. The offset term $C$ is needed to fit the dynamics for times $t> 1\,\text{ns}$. 
The term $\Theta(t)$ is an error function and accounts for the finite width of the probe pulse and the start of excitation at $t = 0$. Applying this model to fit the measurements, we obtain an excellent agreement, see Fig. \ref{fig:Figure3}(b). In the following we use the model to study the amplitude ($A, B$) and frequency ($f_\text{p}$) dependence versus magnetic field and temperature.
\begin{figure}
\includegraphics{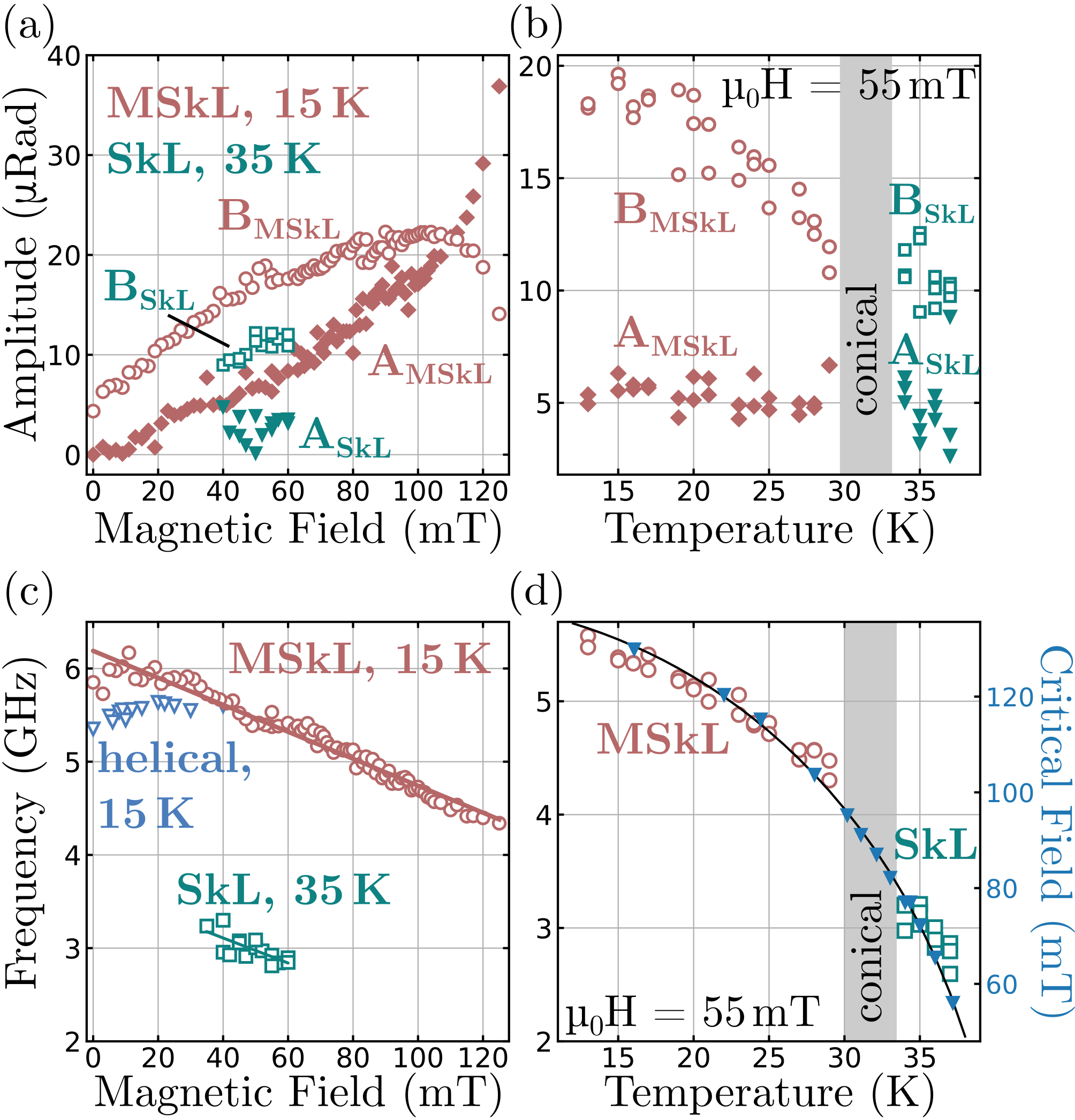}
\caption{\label{fig:Figure4}(a) and (b)	Amplitudes of the exponential ($A$, solid markers) and precessional ($B$, open markers) dynamics as a function of magnetic field and temperature for the MSkL (red markers) and equilibrium SkL (turquoise markers). Fig. (c) and (d) present the magnetic field and temperature dependence of the precession frequency in the MSkL (red dots) and SkL (turquoise square). For comparison the precession frequency as a function of magnetic field of the helical phase at $15\,$K (blue triangle) is plotted. The solid lines in (c) are linear approximations of the data to extract $\Delta f/\Delta B$ and $f_0$. In (d) the critical field $H_{\text{c2}}$ as a function of temperature, extracted from Ref. \cite{bauer2016history}, is plotted and the black line presents a fit of the data to the function ${f_\text{p} = f_0 (1-(T/T_\text{c})^\alpha)^{0.5}}$. In a small temperature gap neither the MSkL, nor the SKL is stable and we observe the conical phase (gray area).}
\end{figure}

We first compare the amplitudes of the precessional and exponential dynamics in the MSkL and equilibrium SkL.
We therefore plot the fit parameter $A$ (exponential) and $B$ (precessional) in Fig. \ref{fig:Figure4}(a) and (b) as function of magnetic field and temperature, respectively. 
In the MSkL, we observe a strongly non-linear magnetic field dependence of $A$ and $B$.
For fields smaller than ${\mu_0 H = 110\,\text{mT}}$ both increase with magnetic field, with the precessional part dominating the spin dynamics. However, for higher fields the amplitude of the precessional mode tends to zero when reaching the phase transition and the exponential dynamics become the leading process. Regarding the equilibrium SkL, the precessional mode strongly dominates the dynamics (${B >> A}$) and scales linearly with the applied magnetic field. In the limited parameter range, the exponential part ($A$) does not show a clear field dependence in the SkL. 
Overall, the TR-MOKE signal (${A+B}$) in the MSkL is much larger compared to the SkL, which can be explained by the increased magnetization at low temperatures. This is supported by the amplitude evolution with temperature (see Fig. \ref{fig:Figure4}(b)), which shows a decrease of the signal with temperature. Here, primarily the precessional dynamics change and the exponential term is rather unaffected. Note, that in a small gap from approximately $30\,\text{K}$ to $34\,\text{K}$ neither the SkL nor the MSkL is observed in our experiment, but the exponential dynamics associated with the conical phase are present. This might be due to an increase of the transient spin temperature above $T_c$ at these base temperatures, such that the MSkL does not recover after laser excitation.

This amplitude study demonstrates that the MSkL offers an extended view on the physical processes contributing to skyrmion dynamics, i.e., collective spin excitation and energy dissipation, compared to the SkL, beneficial for the investigation of generic skyrmion properties.
Furthermore, it shows the higher potential of the MSkL for applications in comparison to its equilibrium state, since larger precessional amplitudes promise larger signal-to-noise ratios. In addition, the increased tunability with the applied magnetic field of the precessional amplitude $B$, but also of the amplitude ratio ($B/A$) in the MSkL are important features for future applications.  

Finally, we focus on the precessional dynamics and compare the frequency $f_{\text{p}}$ of the MSkL and equilibrium SkL. For this purpose the magnetic field and temperature dependence of $f_\text{p}$ are presented in Fig. \ref{fig:Figure4}(c) and \ref{fig:Figure4}(d), respectively. 
We observe a frequency decrease for higher magnetic fields in both states, which is characteristic for the breathing mode of skyrmions \cite{mochizuki2012spin, schwarze2015universal}. Our measurement technique enables the selective detection of the breathing mode, as the other dominating skyrmion eigenmodes, i.e., gyration modes, do not cause a modulation of the average magnetization component out of the sample plane \cite{padmanabhan2019optically}.
Our data reveal a linear dependence of the breathing mode frequency on the magnetic field.
We find a very good qualitative agreement of our phenomenological model with the two models in Ref. \cite{schwarze2015universal, garst2017collective}. Only for magnetic fields approaching the conical transition field the predicted sublinear dependence is not reproduced in our data. Note here, that the helical phase, which is observed for zero-field cooling until $H_{\text{c1}}\approx 40\,\text{mT}$ at $15\,$K, shows a slightly reduced precession frequency and different dispersion compared to MSkL in the same magnetic field range (see blue triangles in Fig. 4c).

By a linear approximation we determine for the SkL and MSkL the frequency change with the applied magnetic field $\Delta f/\Delta B$ and the zero-field frequency $f_0$. For the MSkL we obtain ${\Delta f_{\text{MSkL}}/\Delta B = (-14.5\pm0.4)\,\text{GHz/T}}$ and ${f_{\text{MSkL,0}} = (6.19\pm0.03)\,\text{GHz}}$ and for the equilibrium SkL ${\Delta f_{\text{SkL}}/\Delta B = (-13\pm4)\,\text{GHz/T}}$ and ${f_{\text{SkL,0}} = (3.6\pm0.2)\,\text{GHz}}$, respectively. The frequency change with the applied magnetic field is remarkably similar for the MSkL and equilibrium SkL. This indicates that the characteristics of the dominating skyrmion eigenmodes are preserved in non-thermal equilibrium. As the MSkL is stable in an extended magnetic field range, however, it offers a higher frequency tunability as the equilibrium SkL, beneficial for future applications.

The zero-field frequency $f_\text{0}$ of the equilibrium SkL is much smaller compared to the one in the MSkL, indicating slower spin dynamics in the equilibrium state. In order to investigate this further, we show in Fig. \ref{fig:Figure4}(d) the precession frequency $f_\text{p}$ extracted from the fit of Eq. \ref{eq:phenomenologicalModel} to the experimental data for various base temperatures. 
The precession frequency of the MSkL and equilibrium SkL decreases for higher temperatures
and can be fitted using the phenomenological model ${f_\text{p} = f_0 (1-(T/T_\text{c})^\alpha)^{0.5}}$ with $\alpha = 2.2$ and ${f_0=5.9\,\text{GHz}}$ (black line in Fig. \ref{fig:Figure4}(d)).
The precession frequency change with temperature tracts the $H_{\text{c2}}$ line (see blue markers taken from Ref. \cite{bauer2016history} in Fig. \ref{fig:Figure4}(d)), which describes the temperature dependence of the critical field $H_{\text{c2}}$. This universal scaling of the precession frequency, i.e., the proportionality to the temperature dependence of magnetic order parameters, demonstrates that the observed differences of the skyrmion eigenmodes of MSkL and equilibrium SkL is solely an effect of temperature. Thus, the MSkL is well suited to investigate generic properties of skyrmion dynamics.

In conclusion, this study explores the spin dynamics of the chiral magnet Fe$_{0.75}$Co$_{0.25}$Si in the equilibrium and metastable skyrmion state using time-resolved magneto-optical Kerr effect measurements.
The skyrmion dynamics are well modeled by a phenomenological approach, which separates the spin precession signal, characteristic for the skyrmion breathing mode, from the exponential part, which denotes the energy dissipation. We observe an excellent quantitative agreement of the precession frequency in the equilibrium and metastable skyrmion state considering the universal scaling with temperature.
Our results strongly indicate that the characteristics of skyrmion eigenmodes are preserved far from thermal equilibrium. Besides this fundamental significance, the larger amplitude and enhanced tunability of the MSkL precessional mode point out the higher potential for possible microwave- and magnonics-related applications compared to the equilibrium SkL.

\begin{acknowledgments}
\textbf{Funding:} This work was supported by the European Metrology Research Programme (EMRP) and EMRP participating countries under the European Metrology Programme for Innovation and Research (EMPIR) Project No. 17FUN08-TOPS Metrology for topological spin structures. In part, this study has been funded by the Deutsche Forschungsgemeinschaft (DFG, German ResearchFoundation) under TRR80 (From Electronic Correlations to Functionality, Project No.\ 107745057, Project E1), SPP2137 (Skyrmionics, Project No.\ 403191981, Grant PF393/19 and Grant SCHU 2250/8-1), the excellence cluster MCQST under Germany's Excellence Strategy EXC-2111 (ProjectNo.\ 390814868) and EXC-2123 QuantumFrontiers (ProjectNo. 390837967). Financial support by the European Research Council (ERC) through Advanced Grants No.\ 291079 (TOPFIT) and No.\ 788031 (ExQuiSid) is gratefully acknowledged.
\textbf{Author contributions:} MB and HWS initiated the study. AB and CP grew the crystal. JK performed the TR-MOKE study under the supervision of MB and HF. The data analysis was performed by JK. 
MB, HF, SS, HWS, JK, FGS, AB and CP discussed the results. JK wrote the paper. 
\textbf{Competing interests:} The authors declare that they have no competing interests. 
\textbf{Data availability}: All data needed to evaluate the conclusions of the paper are present in the paper. 
\end{acknowledgments}



\nocite{*}
\bibliography{References}

\end{document}